\documentclass{PoS}

\title{POEMMA: Probe Of Extreme Multi-Messenger Astrophysics}

\ShortTitle{POEMMA}

\author{\speaker{Angela V. Olinto}$^1$\thanks{Corresponding Author}\\
       $^1$Department of Astronomy \& Astrophysics, KICP, EFI, The University of Chicago, Chicago, IL, USA\\
       E-mail: \email{aolinto@uchicago.edu}}
\author{J. H. Adams,$^2$ R. Aloisio,$^3$ L. A. Anchordoqui,$^4$ D. R. Bergman,$^5$ M. E. Bertaina,$^6$ P. Bertone,$^7$ M. Bustamante,$^8$ M. J. Christl,$^7$ S. E. Csorna,$^9$ J. B. Eser,$^{10}$ F. Fenu,$^6$ C. Gu\'epin,$^{11}$ E. A. Hays,$^{12}$ S. Hunter,$^{12}$ E. Judd,$^{13}$ I. Jun,$^{14}$ K. Kotera,$^{11}$ J. F. Krizmanic,$^{12}$ E. Kuznetsov,$^2$ S. Mackovjak,$^{15}$
L. M. Martinez-Sierra,$^{14}$ M. Mastafa,$^2$ J. N. Matthews,$^{5}$ J. McEnery,$^{12}$ J. W. Mitchell,$^{12}$  A. Neronov,$^{16}$ A. N. Otte,$^{17}$ E. Parizot,$^{18}$ T. C. Paul,$^{4}$ J. S. Perkins,$^{12}$ G. Prevot,$^{18}$ P. Reardon,$^{2}$ M. H. Reno,$^{19}$ F. Sarazin,$^{9}$ K. Shinozaki,$^{6}$ F. Stecker,$^{12}$ R. Streitmatter,$^{12}$ T. Venters,$^{12}$  L. Wiencke,$^{10}$ R. M. Young$^{7}$\\
$^2$University of Alabama, Huntsville, AL, USA; $^3$Gran Sasso Science Institute, L'Aquila, Italy; $^4$City University of New York, Lehman College, NY, USA; $^5$University of Utah, Salt Lake City, Utah, USA; $^6$Universita di Torino, Torino, Italy; $^7$NASA Marshall Space Flight Center, Huntsville, AL, USA; $^8$The Ohio State University, Columbus, OH, USA; $^9$Vanderbilt University, Nashville, TN, USA; $^10$Colorado School of Mines, Golden, CO, USA; $^{11}$Institut dÕAstrophysique de Paris, Paris, France; $^{12}$NASA Goddard Space Flight Center, Greenbelt, MD, USA;  $^{13}$Space Sciences Laboratory, University of California, Berkeley, CA, USA; $^{14}$Jet Propulsion Laboratory, Pasadena, CA, USA; $^{15}$Institute of Experimental Physics, SAS, Kosice, Slovakia;
$^{16}$University of Geneva, Geneva, Switzerland; $^{17}$Georgia Institute of Technology, Atlanta, GA, USA; $^{18}$APC-Universite de Paris 7, Paris, France; $^{19}$University of Iowa, Iowa City, IA, USA\\}

\abstract{The Probe Of Extreme Multi-Messenger Astrophysics (POEMMA) mission is being designed to establish charged-particle astronomy with ultra-high energy cosmic rays (UHECRs) and to observe cosmogenic tau neutrinos (CTNs). The study of UHECRs and CTNs from space will yield orders-of-magnitude increase in statistics of observed UHECRs at the highest energies, and the observation of the cosmogenic flux of neutrinos for a range of UHECR models. These observations should solve the long-standing puzzle of the origin of the highest energy particles ever observed, providing a new window onto the most energetic environments and events in the Universe, while studying particle interactions well beyond accelerator energies. The discovery of CTNs will help solve the puzzle of the origin of UHECRs and begin a new field of Astroparticle Physics with the study of neutrino properties at ultra-high energies.}

\FullConference{35th International Cosmic Ray Conference, ICRC217\\
10-20 July, 2017\\
Bexco, Busan, Korea}

\begin{document}

\section{Introduction}

POEMMA, the Probe Of Extreme Multi-Messenger Astrophysics, was selected by NASA for an Astrophysics Probe Mission Concept Study (under ROSES-2016) in early 2017. The comprehensive 18-month POEMMA study involves instrument and mission definition at the Integrated Design Center (IDC) of the Goddard Space Flight Center (GSFC) and an independent cost assessment in preparation for the 2020 Decadal Survey. Here we report on the preliminary concept for POEMMA ahead of the POEMMA Study Report to be submitted to NASA by December 31, 2018. 

\begin{figure}
\begin{center}
\includegraphics [width=1\textwidth]{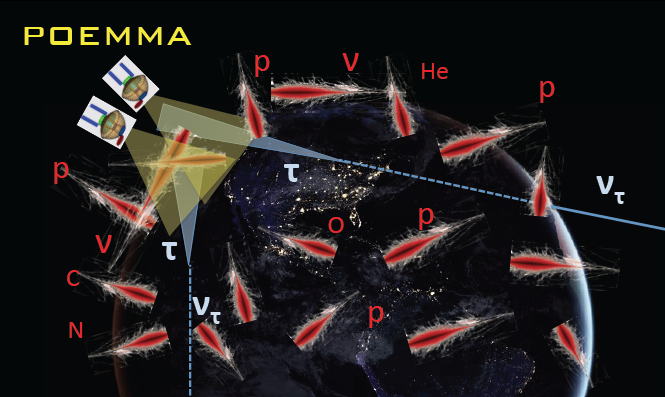}
\caption{POEMMA, the Probe Of Extreme Multi-Messenger Astrophysics, is being designed to observe ultra-high energy cosmic rays (UHECRs) via nitrogen fluorescence of extensive airshowers and cosmogenic tau neutrinos (CTNs) through the Cherenkov radiation of up-going tau lepton decays.}
\label{fig1}
\end{center}
\end{figure}

POEMMA is being designed to enable charged-particle astronomy with a significant increase in exposure to the highest energy particles ever observed, ultra-high energy cosmic rays (UHECRs), and the capability to discover cosmogenic tau neutrinos (CTNs) through the observation of Cherenkov radiation produced by upward-going tau decays. POEMMA will provide an all-sky survey of UHECRs with an order of magnitude larger exposure compared to ground array measurements and two orders of magnitude higher exposure in fluorescence mode when compared to ground fluorescence observatories (significantly improving the determination of composition above 10s of EeVs). The increase in exposure combined with the full-sky coverage should reveal the sources of these extremely energetic particles that are known to reach Earth from extragalactic sources and that are yet to be identified. These unidentified sources achieve extreme acceleration  through mechanisms that are not presently understood. 

As UHECRs propagate from distant extragalactic sources they interact with cosmic background radiation losing energy through the Greisen-Zatsepin-Kuzmin (GZK) effect~\cite{GZK} and producing cosmogenic neutrinos~\cite{BZnu}. Observations from the leading UHECR observatories, the Pierre Auger Observatory~\cite{Auger} in Mendoza, Argentina, and the Telescope Array (TA)~\cite{TA} in Utah, USA, show a spectral shape consistent with the GZK effect, but also explainable by the maximum energy of the unidentified astrophysical accelerators, $E_{\rm max}$. Higher statistics measurements of both the flux and the composition of UHECRs  above 10 EeV (1 EeV = $10^{18}$ eV) together with the detection of the flux of cosmogenic neutrinos can settle this long-standing mystery (see e.g.,~\cite{review} for more details). POEMMA is being designed for a significant increase in statistics and the detection of Cherenkov radiation from up-going tau decays produced by cosmogenic neutrinos. The observation of cosmogenic neutrinos will help solve the puzzle of the origin of UHECRs and begin a new field of Astroparticle Physics with the study of neutrino properties at energies orders of magnitude above those reached by human-made accelerators.  

\begin{figure}
\begin{center}
\includegraphics [width = 1\textwidth]{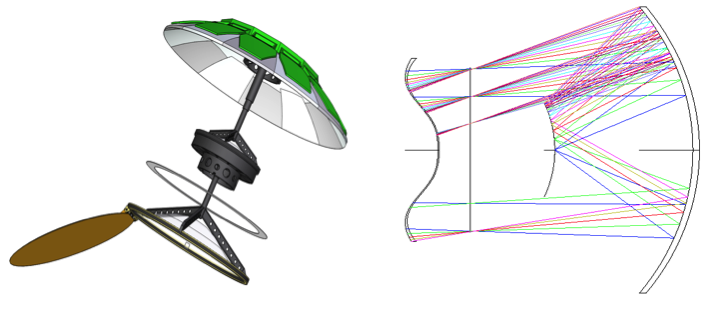}
\caption{POEMMA mirrors: Right - optical design for a f/0.77 Schmidt mirror and corrector lens, with 45$^{\rm o}$ field of view (FoV); Left - POEMMA mirror and focal surface and shutter deployed.}
\label{fig2}
\end{center}
\end{figure}

The POEMMA design combines the concept developed for the Orbiting Wide-field Light-collectors (OWL)~\cite{OWL1,OWL2} mission, the experience of the Extreme Universe Space Observatory (EUSO) on the Japanese Experiment Module (JEM-EUSO)~\cite{JE_ExpAst}  fluorescence detection camera as recently flown on EUSO-SPB1 by a NASA Super Pressure Balloon (SPB)~\cite{EUSO_SPB1} from Wanaka, New Zealand, with the recently proposed CHerenkov from Astrophysical Neutrinos Telescope  (CHANT)~\cite{CHANT} concept to form a multi-messenger probe of the most extreme environments in the Universe. 

Building on the OWL concept, POEMMA is composed of two identical satellites flying in formation with the ability to observe overlapping regions during moonless nights at angles ranging from Nadir to just above the limb of the Earth. The satellites' altitude is planned to be varied in tandem from about 525 km up to 1,000 km with different separations and pointing strategies. 

POEMMA satellites detect UHECRs through the observation of particle cascades (or extensive airshowers) produced by the interaction of UHECRs with the Earth's atmosphere. Particles in extensive airshowers excite nitrogen molecules in the atmosphere, which fluoresce in the ultraviolet (UV) and can be observed by ultra-fast UV cameras. The fluorescence technique has been perfected by the leading ground-based UHECR observatories, Auger and TA, while EUSO-Balloon~\cite{EUSO-Balloon} and EUSO-SPB1~\cite{EUSO_SPB1}  recently pioneered the fluorescence technique from suborbital space.

Each POEMMA satellite consists of a large Schmidt telescope with a deployable mirror similar to the OWL concept of a 7-meter diameter deployable optics system. To reach a lower energy threshold, the POEMMA Schmidt telescope is planned to have an f-number of 0.77, which leads to a 6.5 diameter mirror with an optical aperture of about 14 m$^2$.  This is approximately 2 times the aperture for OWL (7.07 m$^2$), which was an f/1 system. Each POEMMA telescope monitors a 45$^{\rm o}$ field of view (FoV) and a 2.3 m diameter optical aperture with a single corrector plate. A lens-cap lid and a ``jiffy-pop'' cover protect the mirror of stray light and micrometeoroid. The mirrors act as large light collectors with modest imaging requirements.

\begin{figure}
\begin{center}
\includegraphics [width= .60\textwidth]{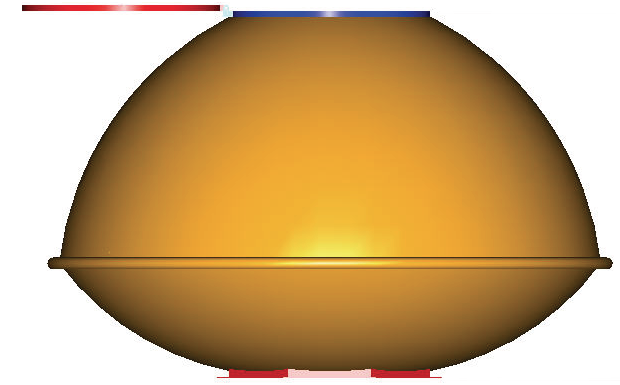}
\caption{POEMMA Mirror Cover: a ``jiffy-pop'' cover (in yellow) protects the mirror of stray light and micrometeoroids,  a moving lens cap (in red) opens during observing runs and closes to protect the corrector lens (in blue) when telescopes are not taking data.}
\label{fig3}
\end{center}
\end{figure}

The POEMMA focal surface is composed of a hybrid of two types of cameras: about 90\% of the focal surface is dedicated to the POEMMA fluorescence camera (PFC), while POEMMA Cherenkov camera (PCC) occupies the crescent moon shaped edge of the focal surface which images the limb of the Earth. The PFC is composed of EUSO Photo Detector Modules (PDM) based on multi-anode photomultiplier tubes (MAPMTs) as flown in sub-orbital space in EUSO-Balloon~\cite{EUSO-Balloon} and EUSO-SPB1~\cite{EUSO_SPB1} and soon to be deployed in the International Space Station (ISS) as mini-EUSO~\cite{mini_EUSO}. The typical time between images for the PFC is about 1 $\mu$sec.  The much faster POEMMA Cherenkov camera (PCC) is composed of Silicon photo-multipliers (SiPMs) with sampling time of about 100 nsec. Note that SiPMs flew on EUSO-SPB1 and soon to be tested in space with mini-EUSO. The PFC registers UHECR tracks from Nadir to just below the Earth's limb, while the PCC registers light within the Cherenkov emission cone of up-going showers around the limb of the Earth and also from high energy cosmic rays above the limb of the Earth. 

\begin{figure}
\begin{center}
\includegraphics [width= 1\textwidth]{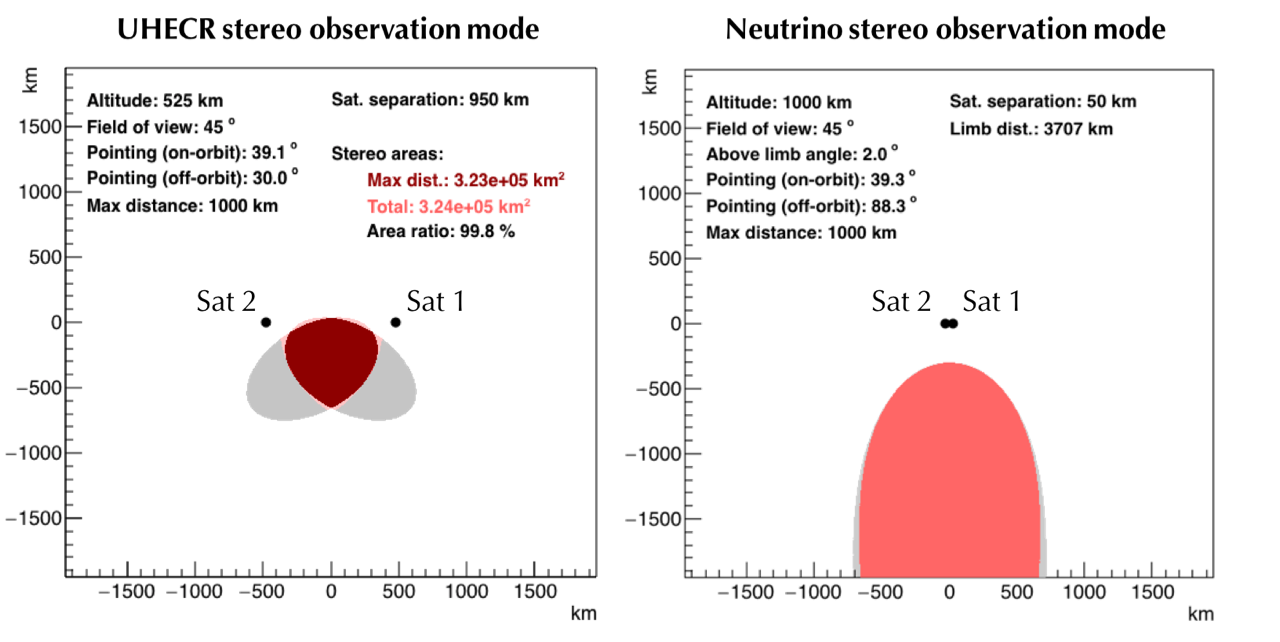}
\caption{Examples of POEMMA stereo observation modes. Left: Fluorescence Stereo Mode with satellites altitude of 525 km, separation between satellites 950 km, Stereo area 3.24 $\times 10^5 \ {\rm km}^2$; Right: Cherenkov Stereo Mode with satellites altitude of 1000 km, separation between satellites 50 km, limb observations at distance 3,707 km. }
\label{fig3}
\end{center}
\end{figure}

\section{Science Goals} 

POEMMA will provide a new window on the Universe's most energetic environments and events. POEMMA is being designed to significantly increase the statistics of observations of different components of UHECRs and neutrino species over a wide range of energy with greater focus on the highest energies ever observed . The instrument design will focus on answering following science question

{\it What objects can accelerate particles to ultra-high energies?} 
To discover the sources of UHECRs, POEMMA will survey from space two orders of magnitude larger volumes of the atmosphere when compared to ground observatories over the full sky with nearly uniform exposure. The related questions of {\it how are the sources distributed in the sky?} will be addressed with a full sky map of UHECRs with significantly higher statistics at the highest energies, where pointing to the sources becomes feasible (above $\sim$10 EeV). POEMMA is designed to reach unprecedented geometrical apertures $>10^6 \ {\rm km}^2$ sr yr, which, after duty cycle corrections, correspond to annual exposures of more than $10^5$ km sr yr at the highest energies. POEMMA will also have high angular resolution ($\sim 1^{\rm o}$).

POEMMA will enable far more sensitive sky maps leading to the discovery of the brightest sources of UHECRs in the sky, which are likely to be relatively nearby (within $\sim$100 Mpc). The appearance of nearby sources in the sky is regulated by the GZK effect, which suppress the contribution from very distant accelerators at energies above 10s of EeV. The EeV UHECR sky is isotropic because sources throughout the observable Universe contribute without any damping, while the 100 EeV UHECR sky should only show the nearby sources as the GZK effect obfuscates sources further then 100 Mpc moving closer to 10 Mpc at 100 EeV. A clear source distribution will become apparent when a high statistics map above 60 EeV is produced by POEMMA. In addition, observations above 10s of EeV  avoid large deviations (compared to the few degrees angular resolution) from source to arrival directions on Earth due to cosmic magnetic fields. The angular size of the nearby sources in the sky will probe the magnitude and structure of extragalactic and galactic magnetic fields.  Above 10s of EeV, {\it Charged-Particle Astronomy} is finally attainable. 

\begin{figure}
\begin{center}
\includegraphics [width= .90\textwidth]{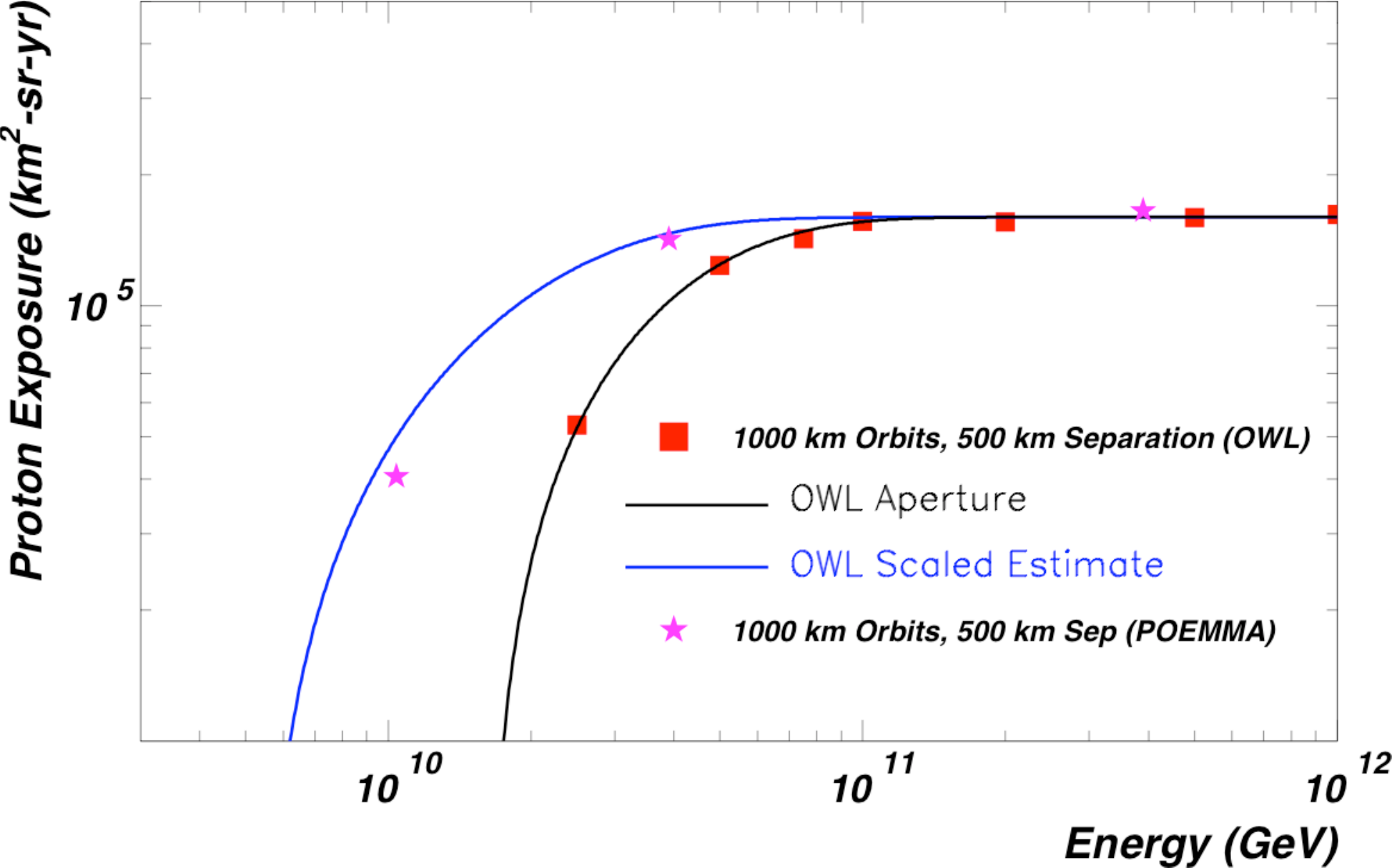}
\caption{Preliminary POEMMA  yearly exposure to protons assuming a 10\% duty cycle. Estimates and simulations based on the scaled version of the OWL design. Final exposure to be determined at the conclusion of the POEMMA concept study.}
\label{fig3}
\end{center}
\end{figure}

POEMMA will also address {\it what is the composition of the UHECRs above 10 EeV and how does it evolve as energies reach 100 EeV?} POEMMA stereo observations of UHECRs will yield significant increase in measurements of the maximum of extensive airshowers, $X_{\rm max}$, with rms resolution of at least $\sim$ 60 g/cm$^2$ and a large enough sample of well reconstructed events with better $X_{\rm max}$ separation to distinguish light and heavier nuclei above 10 EeV. These composition measurements together with spectrum and sky distribution of anisotropies will determine the source class of UHECRs. 

{\it What is the flux of cosmogenic neutrinos?} By observing the Cherenkov signal from tau decays from the limb of the Earth, POEMMA will determine the flux level of cosmogenic neutrinos for a wide range of UHECR source models. The cosmogenic neutrino flux is very sensitive to the different candidate UHECR sources and add an extra dimension in determining these unidentified sources. In addition, POEMMA can observe tau neutrinos at lower energies addressing the additional question of: {\it What is the flux of astrophysical tau neutrinos?}.

POEMMA will search for astrophysical and cosmogenic neutrinos with two techniques. With the same system designed to observe UHECRs, the PFC  can detect deeply penetrating horizontal showers initiated by all flavors of EeV neutrinos in the atmosphere. In addition, the PCC based on the CHANT concept can observe the signal produced from tau neutrinos from 10 PeV (where astrophysical IceCube neutrinos are expected) to 10 EeV (where cosmogenic neutrinos can be discovered). The cosmogenic neutrino flux is a by-product of the propagation of UHECR through the GZK interactions via neutron and muon decays. The spectrum of cosmogenic neutrinos depends on the composition and source distribution of UHECRs. A main peak is generically produced around 1 to 10 EeV. A secondary peak, around PeV energies, may occur in models where protons dominate the UHECR composition due to neutron decay. The PeV peak is generally subdominant to the astrophysical neutrinos observed by IceCube, thus the most important evidence of the GZK process is the detection of neutrinos in the 1 to 10 EeV energy range. This range will establish a new Astroparticle Physics field with the study of neutrino properties at EeV energies, well above energies accessible in the laboratory.

POEMMA will also be sensitive to ultra-high-energy photons (UHEP) in the most optimistic astrophysical scenarios. The UHEP flux is highly dependent  on the model of UHECR sources, being highly sensitive to the location of the closest sources. UHEP are the dominant component of models based on relic decays from the early universe, including super-heavy dark matter. A clear detection of UHEPs would be momentous discovery.

Additional science themes include the study of {\it how strong are magnetic fields in the extragalactic medium?} Cosmic magnetic fields are traditionally challenging to measure and very little is known about magnetic fields outside galaxies and clusters of galaxies. The pointing pattern to UHECR sources will constrain these extragalactic fields directly. POEMMA will also study atmospheric phenomena in the optical and the UV such as transient luminous events in the upper atmosphere, will observe meteors arriving on Earth, and will search for meteorite (see, e.g., similar studies for JEM-EUSO in~\cite{TLE,Meteor}).

\section{Calibration and Atmospheric Monitoring}

POEMMA observes UHECRs over a very large area (approximately the size of the state of Utah, USA) moving across the globe at  the orbital speed of $\sim$7 km/sec. The observed volume of the atmosphere  will include variable amounts of clouds with altitudes from sea level to $\sim$15 km and variable boundary layer aerosols. Consistency between the stereo views of an event provides a powerful tool for understanding scattering through intervening high clouds or aerosol layers. 

A monitoring and calibration system is being designed for POEMMA, which includes a steerable UV laser, an infrared camera, a LIDAR, and a system of ground-based UV LEDs and lasers forming a worldwide network to continuously calibrate the triggering, intrinsic luminosity, and pointing accuracy of POEMMA's observations.

\section{Mission Overview} 

The POEMMA mission involves two satellites flying in formation in a relatively low-altitude, near-equatorial orbit, with each satellite operating independent. Once in orbit, the satellites will deploy from their stowed, launch configuration.  The large mirror, focal surface, and corrector plate will be deployed along with the solar array, sun shield, and the antennas. 

One of the key items of the POEMMA instrument and mission definition at the IDC is to determine if both satellites can be launched as a dual-manifest on the same launch vehicle, as was considered for OWL. The satellites will be inserted into a circular orbit at an inclination of about 28.5$^{\rm o}$ and an initial altitude of 525 km moving further to 1000 km as initial science goals are completed. The satellites will maneuver to the desired separation distance and attitude. To search for Cherenkov signals from Earth-skimming neutrinos the satellites will be separated such that both fall within the same Cherenkov light pool and oriented to view the same area on the limb of the Earth. To focus on extreme energy cosmic rays, the separation will  be larger and both satellites will be oriented to view the same area near the nadir. A sequence of observing formation stages will be developed to address each science goal for the minimum 3 year mission with a 5 year mission goal.

\section{Acknowledgement}
The POEMMA concept study is funded by NASA Award NNX17AJ82 at the University of Chicago and GSFC.

\end{document}